\documentclass[prl,aps,twocolumn,pdffig]{revtex4}
\usepackage{graphicx}
\usepackage{dcolumn}
\usepackage{bm}
\makeatletter

\newcommand{\Rmnum}[1]{\expandafter\@slowromancap\romannumeral #1@}
\makeatother
\begin{document}
\title {$1/f$ noise as a probe for investigating band structure in graphene}
\author{Atindra Nath Pal,$^1$ Subhamoy Ghatak,$^1$ Vidya Kochat,$^1$ Sneha
E. S.,$^1$ Arjun B. S.,$^2$ Srinivasan Raghavan,$^2$ and Arindam Ghosh$^1$} \vspace{1.5cm}
\address{$^1$ Department of Physics, Indian Institute of Science, Bangalore 560 012, India}
\address{$^2$ Materials Research Center, Indian Institute of Science, Bangalore 560 012, India}

\begin{abstract}
A distinctive feature of single layer graphene is the linearly dispersive energy bands, which in case of multilayer graphene become parabolic.
Other than the quantum Hall effect, this distinction has been hard to capture in electron transport. Carrier mobility of graphene has been
scrutinized, but many parallel scattering mechanisms often obscure its sensitivity to band structure. The flicker noise in graphene depends
explicitly on its ability to screen local potential fluctuations. Here we show that the flicker noise is a sensitive probe to the band structure
of graphene that vary differently with the carrier density for the linear and parabolic bands. Using devices of different genre, we find this
difference to be robust against disorder in the presence or absence of substrate. Our results reveal the microscopic mechanism of noise in
Graphene Field Effect Transistors (GraFET), and outline a simple portable method to separate the single from multi layered devices.
\end{abstract}


\maketitle


In a field-effect device, the flicker noise manifests in slow fluctuations in the drain-source current due to the fluctuations in the channel
conductivity, $\sigma$. The flicker noise is often called the $1/f$-noise because of its power spectral density, $S_\sigma(f) \propto 1/f^\nu$,
where $f$ is frequency, and $\nu \approx 1$. The $1/f$-noise has been studied extensively in metal-oxide field effect transistors (MOSFET),
where the trapping and detrapping of charge at the channel-oxide interface represent a collection of two-state fluctuators~\cite{correlated
model_jayaraman}. A wide distribution in the fluctuator switching rate leads to $1/f$ noise in these devices. Similar mechanism of noise have
been assumed for carbon nanotube field-effect devices as well~\cite{CNT_avouris}, where the trapping events close to the nanotube-metallic lead
Schottky barriers cause fluctuations in the effective gate voltage~\cite{CNT_charge noise}.

\begin{figure}[!t]
\begin{center}
\includegraphics [width=1\linewidth]{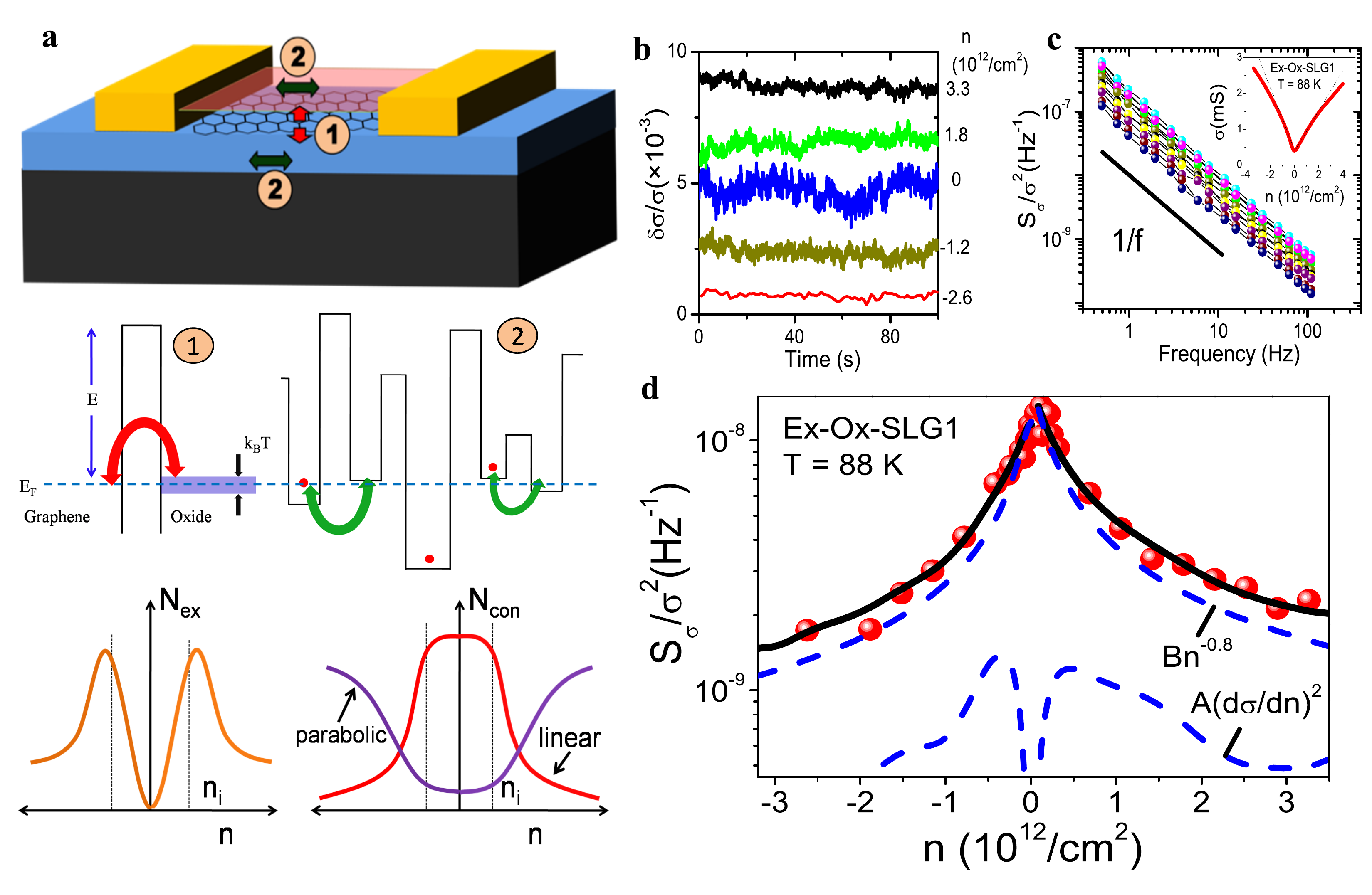}
\end{center}
\caption{(a) Schematic of the GraFET device showing two different charge noise mechanisms. Process (1) corresponds to the exchange noise due to
the charge transfer between graphene and its environment. Process (2) depicts the configuration noise arising from the rearrangement of trapped
charges within the environment. The density dependence of each component has been schematically shown at the bottom depicting opposite nature of
configuration noise for linear and parabolic bands. (b) Time domain conductivity fluctuations at different carrier densities. (c) Typical noise
power spectra $S_{\sigma}/{\sigma}^2$ at various back gate voltages, showing $1/f$ characteristics. Inset shows conductivity ($\sigma$) vs.
density ($n$) for a substrated SLG device with dotted line indicating the linear region. (d) $S_{\sigma}/{\sigma}^2$ (at 1 Hz) vs. density ($n$)
data for a substrated SLG device, fitted with Eq.~2 (see text). The contributions from the two different noise mechanisms are shown by the
dashed lines.}
 \label{figure1}
 \end{figure}

In spite of the close architectural similarity to MOSFETs or nanotube-based FETs, the microscopic understanding of noise in GraFETs is rather
limited~\cite{Avouris 1,{atin PRL},{atin APL},{Charge noise},{Noise_Xu},{Noise_Balandin},{Noise_suspended}}. An augmented charge noise
model~\cite{charge_noise_IBM}, developed in the context of carbon nanotubes, fails to describe the carrier density ($n$) dependence of noise in
most single layer graphene (SLG) devices close to charge neutrality (the Dirac point)~\cite{Avouris 1,{atin APL}}. The model does not consider
the graphene band structure explicitly, and hence fails to account for distinct noise behavior in SLG and bilayer (BLG) GraFET
devices~\cite{Avouris 1,{atin PRL}}, which becomes more dramatic at lower temperatures ($T$). The influence of quenched disorder, and related
charge inhomogeneity~\cite{martin}, which are both very serious technological bottlenecks, are also not known. With emerging techniques of
realizing graphene at commercial production scale, such as chemical vapor deposition on metals~\cite{CVD_science} or epitaxial growth on
SiC~\cite{SiC_Science1}, the behavior of noise at low carrier density, $|n| < n_i$, where $n_i$ is the scale below which charge distribution in
graphene becomes inhomogeneous, needs to be addressed. In this work we explore the possibility of a global framework within which flicker noise
in GraFET devices of different genre~\cite{CVD_science,{SiC_Science1},{GNR},{suspended graphene}} and substrates~\cite{high
K,{graphene_BN},{mica nature}} can be understood and analyzed.

We consider the generic GraFET structure in the schematic of Fig.~1a, where the local environment of the graphene film consists of the
underlying insulating substrate, and surface adsorbates (or top gate dielectric, if any). Focussing on the noise that arises due to fluctuating
charge distribution (FCD) around the graphene film, two processes are identified: (1) Exchange of charge between graphene and its environment,
for example through trapping-detrapping process, which involves time-dependent changes in $n$. This causes a charge exchange noise, $N_{ex}
\propto (d\sigma/dn)^2$ through correlated number and mobility ($\mu$) fluctuations~\cite{correlated model_jayaraman}. (2) The second process
consists of a slow rearrangement of charge within the local environment of graphene, for example, random migration of trapped charges within the
substrate or surface adsorbates (Process 2 in the schematic in Fig.~1a), and referred as configuration noise ($N_{con}$). This process alters
the disorder landscape due to Coulomb potential from trapped charges leading to random fluctuations in the scattering cross-section
($\Lambda_c$). Within a ``local interference'' framework~\cite{Pelz}, $\delta\Lambda_c \sim \Lambda_c \propto |v_q|^2$, where $v_q$ is the
screened Coulomb potential of the trapped charge. Thus at large $n$, $N_{con} \propto \ell^2 |v_q|^4 \sim |n|^\gamma$, where $\ell$ is the mean
scattering length~\cite{Pelz}, where $\gamma$ is determined by the $n$-dependence of $\ell$ and $v_q$, and hence is sensitive to the nature of
graphene band structure. Assuming Boltzman transport in Thomas-Fermi screening, the total normalized noise power spectral density can be written
as (see supplementary material for full derivation),

\begin{eqnarray}
\label{eq2} S_\sigma(f)/\sigma^2 = A(T)(d\sigma/dn)^2 + B(T){\cal N}_c(n)
\end{eqnarray}

\noindent To the leading order the parameters $A(T)$ and $B(T)$ are independent of $n$ irrespective of the band structure, but depend on $T$.
The function ${\cal N}_c(n) = |n|^\gamma$ for $|n| \geq n_i$, while for $|n| \leq n_i$, the charge distribution in graphene disintegrates into
electron-hole puddles~\cite{martin}, and noise is determined by changes in the percolation network and weak links between puddles. In this
regime, we take ${\cal N}_c(n) \rightarrow$ constant (see schematic in Fig.~1a). In analyzing the noise data, the parameters $A, B, n_i$ and
$\gamma$ were kept as fitting parameters, and $d\sigma/dn$ was obtained from the $\sigma-n$ data. Due to particle-hole asymmetry~\cite{gordon}
the electron and hole-doped regimes were fitted separately~\cite{phonon_avouris}.

\begin{figure}[!t]
\begin{center}
\includegraphics [width=1\linewidth]{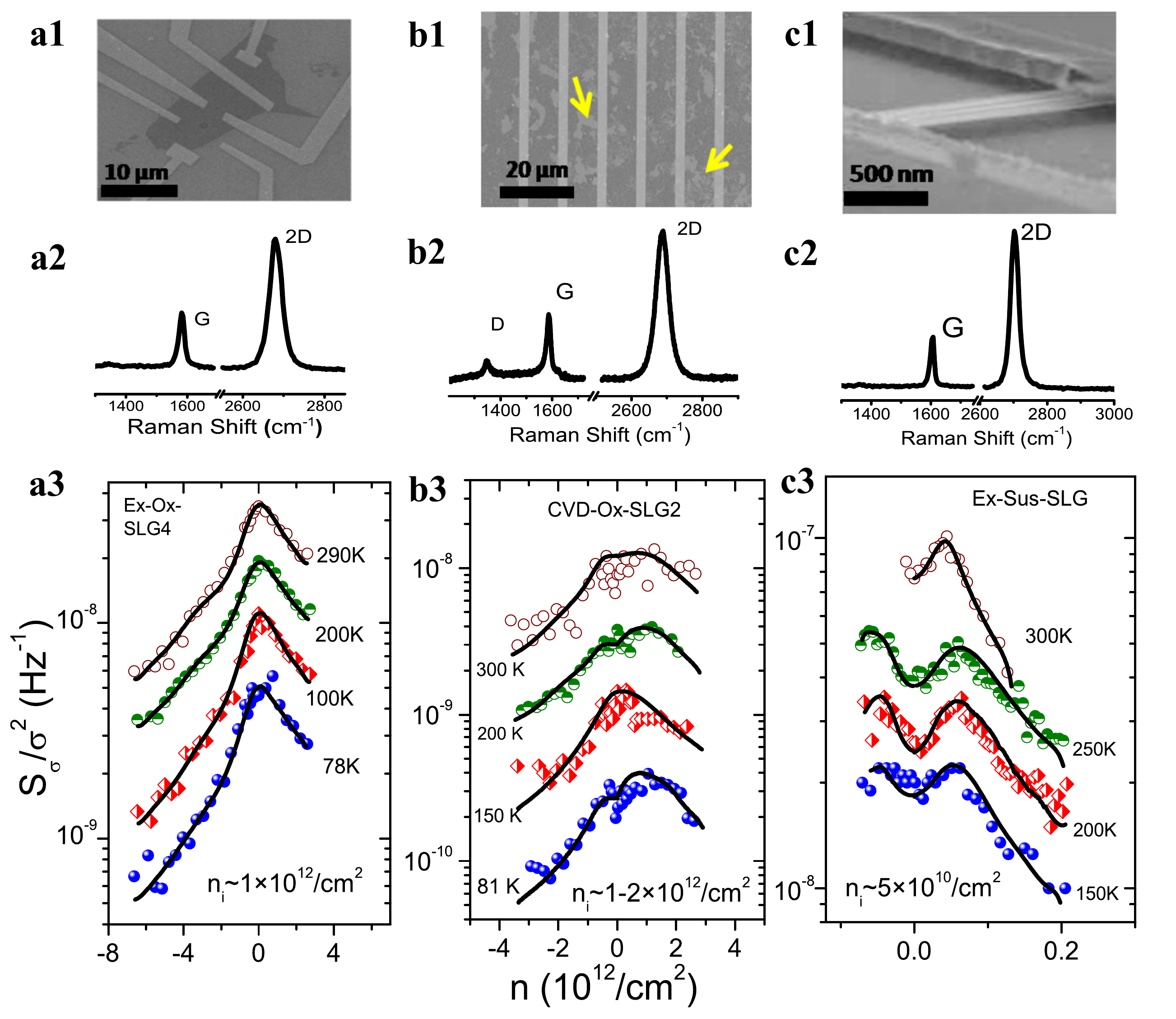}
\end{center}
\caption{The noise magnitude, $S_{\sigma}/{\sigma}^2$ (at 1 Hz) vs. carrier density ($n$) at different temperatures, fitted with the FCD model
for substrated SLG, CVD-graphene, and suspended SLG in Fig.~2a3-c3 respectively. The top panel (a1-c1) shows the SEM images of the devices, and
corresponding Raman spectrum for each device is shown in Fig.~a2-c2. For clarity, the noise traces at different temperatures have been shifted
vertically. The arrows in Fig.~2b2 indicate typical voids and ruptures in CVD graphene.} \label{figure2}
\end{figure}

We have used seven different types of graphene field effect devices in our experiments which include exfoliated single and multilayer graphene
on oxide substrate, freely suspended single layer graphene, and chemical vapor deposition (CVD)-grown graphene on Si$_2$ (see Table~\ref{table1}
for details). Graphene flakes were prepared on $300$~nm SiO$_2$ on $n^{++}$ doped silicon substrate (the backgate) by micromechanical
exfoliation of natural graphite (NGS Naturgraphit GmbH). In all the cases, we have used RCA cleaning of the substrate, and standard electron
beam lithography technique followed by thermal evaporation of 40-50 nm gold ($99.99\%$) to fabricate the devices. For suspended graphene
transistor, $100$~nm thick gold electrodes were made, followed by etching of the underneath oxide by $1:6$ buffered HF solution for
$2.5$~minutes. Finally, the devices were released in a critical point dryer. No current annealing or Ar/H$_2$ annealing was used in our
experiments to remove the acrylic residues in any of the devices. To avoid large electrostatic force, only small range of gate voltage was
scanned, corresponding to $n \lesssim 2\times10^{11}$~cm$^{-2}$. The CVD-graphene was grown by thermal decomposition of methane on $25~\mu$m
thick copper foil at $1000^{\circ}$C. Methane was introduced into the chamber at a rate of $35$ sccm and a pressure of $4$ Torr for a growth
time of $8$ minutes after which the chamber was cooled down to room temperature. The susbsequent processes involved PMMA coating, dissolving
copper foil with Ferric Chloride ($1.75$~g FeCl$_3$/$5$ ml conc. HCl/ $50$~ml de-ionized water), transfer onto the Si/SiO$_2$ substrate, coating
a second PMMA layer, and finally cleaning with acetone/IPA. Noise in the graphene devices were measured in low-frequency ac four- and two-probe
methods in a high vacuum environment. See Ref~\cite{arindam_arxiv} for details. The excitation was below 50 $\mu$A to avoid heating and other
non-linearities, and verified by quadratic excitation dependence of voltage/current noise at a fixed resistance $R$. The background noise was
measured simultaneously, and subtracted from the total noise.

The FCD model was first examined with an exfoliated SLG device (Ex-Ox-SLG1). The inset of Fig.~1c shows $\sigma$ to vary linearly at low $|n|$,
indicating scattering from charged impurities. The normalized fluctuations, $\delta\sigma/\sigma$, peak at the Dirac point, as evident from the
time traces in Fig.~1b. We find $S_\sigma(f) \propto 1/f$ at all $n$ (Fig.~1c). Both four- and two-probe measurements yielded same results,
indicating negligible contribution from contact noise. $S_\sigma/\sigma^2$ decreases monotonically as $|n|$ is increased on both electron and
hole doped sides (Fig.~1d). Fitting Eq.~\ref{eq2}, shown in solid dark line, yields excellent agreement. Two key factors can be noted here.
First, $N_{con}$ exceeds $N_{ex}$ at all $n$, particularly at lower $|n|$ (dashed lines in Fig.~1d). Indeed, the peak in $S_\sigma/\sigma^2$ at
the Dirac point can be attributed to larger configurational noise, {\it i.e.} enhanced sensitivity of graphene to alteration in disorder
landscape, with charge noise being minimal since $\sigma$ varies weakly with $n$ in this regime. Second, the fit yields $\gamma \sim -1.0$,
which was found to be generic to other SLG devices as well (see also Fig.~4a). Within the Thomas-Fermi approximation, $v_q \sim 1/k_F \sim
1/\sqrt{|n|}$ for SLG ($k_F$ is the Fermi wave vector)~\cite{sdsarma_BLG}. Taking $\ell \propto |n|^\epsilon$, where $\epsilon \simeq 0.5$ for
both screened Coulomb~\cite{sdsarma_BLG} and interface phonon scattering~\cite{phonon_natnano,{phonon_avouris}} in graphene, we get $N_{con}
\sim \ell^2|v_q|^4 \sim |n|^{-1}$, as indeed observed.

\begin{figure}[!tbp]
\begin{center}
\includegraphics [width=1\linewidth]{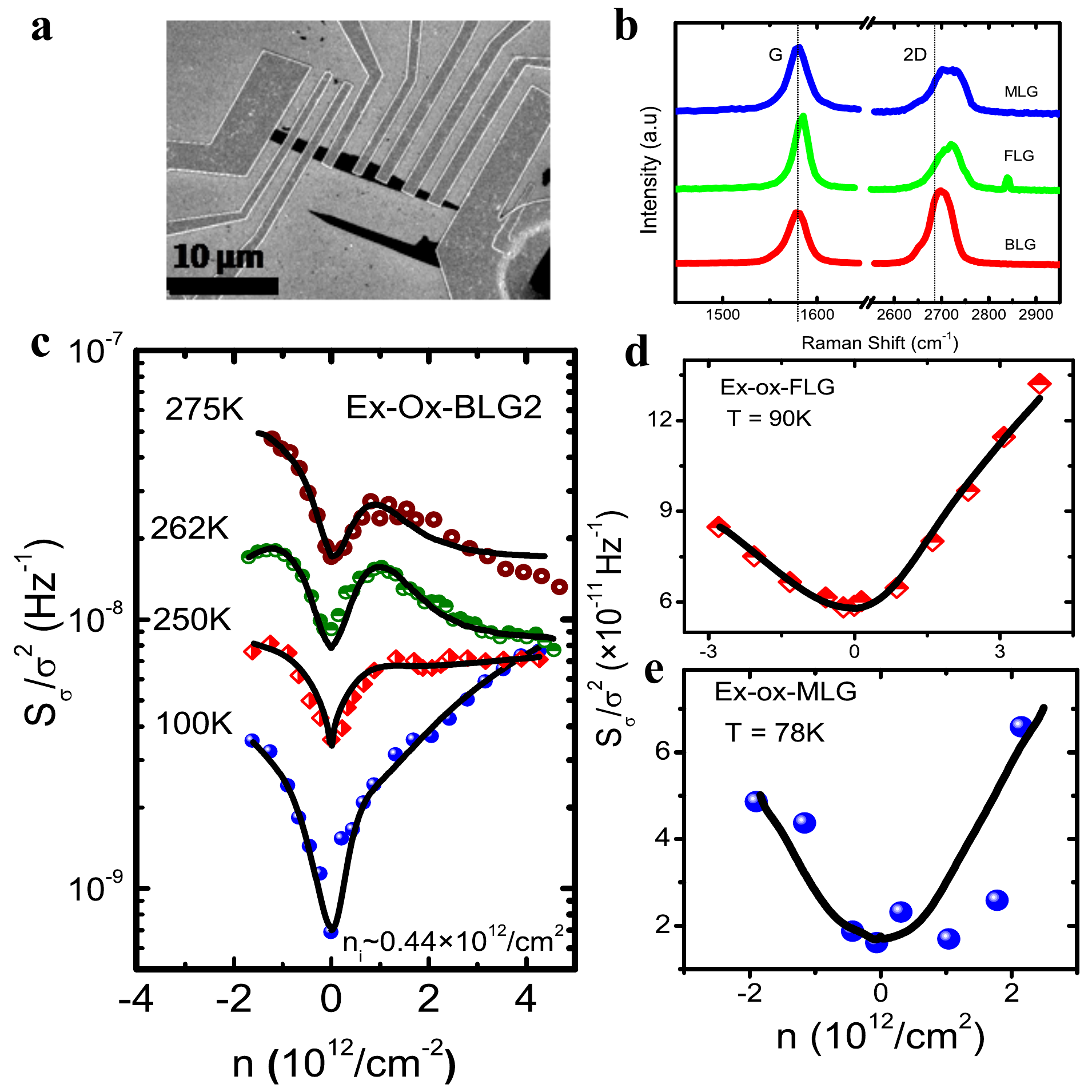}
\end{center}
\caption{(a) SEM micrograph of the BLG device used in the experiment. (b) Raman spectra for BLG, FLG and MLG showing the characteristic G and 2D
peaks. (c) The noise magnitude, $S_{\sigma}/{\sigma}^2$ (at 1 Hz) vs. carrier density ($n$) at different temperatures, fitted with the FCD model
for the BLG device, while, similar plots have been shown for FLG (T = 90K) and MLG (T = 78K) devices in
 3d \& 3e respectively. } \label{figure3}
\end{figure}

The validity of the analysis was then confirmed to be robust with SLG against varying levels of quenched disorder. Here, the noise in three SLG
devices obtained by, (1) mechanical exfoliation on SiO$_2$ substrate (Ex-Ox-SLG4, Fig.~2a1-a3), (2) chemical vapor deposition (Cvd-Ox-SLG2,
Fig.~2b1-b3), and (3) mechanically exfoliated suspended single layer graphene (Ex-Sus-SLG, Fig.~2c1-c3), are compared. For Cvd-Ox-SLG2, the
transfer process onto the SiO$_2$/Si substrate introduce considerable disorder, both structural (ruptures/voids) and foreign charged/uncharged
residues (see the electron micrograph and the disorder peak in the Raman spectrum in Fig.~2b2), resulting in rather poor $\mu \sim
400$~cm$^2$/Vs (see Table~I). In spite of this, we find overall similarity in the behavior of $S_\sigma/\sigma^2$ in all SLG devices on oxide
substrate (including Ex-Ox-SLG1 in Fig.~1d). The magnitude of $n_i$ ($\approx 1 - 2\times 10^{12}$~cm$^{-2}$) in Cvd-Ox-SLG2 is significantly
larger than typical $n_i$ in exfoliated SLG devices, confirming a greater inhomogeniety in the former. In all cases Eq.~\ref{eq2} provides an
excellent fit to the data with $\gamma \approx -1$ (see Fig.~4a). For Cvd-Ox-SLG2, significant electron-hole asymmetry due to impurity
scattering leads to asymmetric noise behavior. It is remarkable that in spite of higher level of structural disorder, noise in CVD-grown
graphene continues to be dominated by fluctuating charge distribution, indicating that migration of structural disorder is mostly frozen well up
to the room temperature.

The noise behavior is qualitatively different in the low disorder limit, as indicated by the suspended SLG device Ex-Sus-SLG (Fig.~2c1-c3).
Measurements were performed on as-fabricated device, which was mildly electron doped due to surface residues and contamination (see Table~I).
The noise magnitude $S_\sigma/\sigma^2$ at $f = 1$~Hz is shown in Fig.~2c3, which was considerably lower than the other SLG devices when
normalized by device size and low operating $n$ (see also Fig.~4b). Intriguingly, the noise in Ex-Sus-SLG varies nonmonotonically with $n$ at
all $T$. In spite of this difference, fitting Eq.~\ref{eq2} gives excellent agreement, albeit with a much greater contribution from $N_{ex}
\propto (d\sigma/dn)^2$ which results in the non monotonic behavior. We believe the observed noise in our suspended graphene to be due to
residual surface contamination, in particular the residues of the electron beam resist (PMMA). The analysis also confirms $\gamma$ to be
negative which approaches $\sim -2.0$ near room temperature (Fig.~4a) due to a nearly constant $\ell$ from competing scattering processes.

\begin{table*}
\caption{\label{table1}Details of the devices .}
\begin{ruledtabular}
\begin{tabular}{cccccc}
 Device & Growth &Layer & Substrate & Device area (L$\times$W)\footnote{both dimensions in $\mu$m} & Mobility \footnote{in cm$^2$/V.s}\\ \hline
 Ex-Ox-SLG1 & Exfoliation & 1 & SiO$_2$ & $4.5 \times 3.5$ & $8000$ \\
 Ex-Ox-SLG4 & Exfoliation & 1 & SiO$_2$ & $3.1 \times 5$ & $3500$  \\
 Ex-Ox-BLG2 & Exfoliation & 2 & SiO$_2$ & $1.2 \times 5.2$ & $1200$  \\
 Cvd-Ox-SLG2 & CVD & 1 & SiO$_2$ & $15 \times 60$ & $400$ \\
 Ex-Sus-SLG & Exfoliation & 1 & Suspended & $1.5 \times 2$ &$20,000$ \\
 Ex-Ox-FLG & Exfoliation & 3-4 & SiO$_2$ & $2 \times 3$ & $2450$  \\
 Ex-Ox-MLG & Exfoliation & 14-15 & SiO$_2$ & $1.9 \times 6$ & $1200$  \\
\end{tabular}
\end{ruledtabular}
\end{table*}

Noise measurements on BLG led to a rather striking result. Fig.~3a-c show a micrograph, Raman spectrum, and $n$-dependence of noise in an
exfoliated BLG device (Ex-Ox-BLG2) on identically treated SiO$_2$/Si substrates. No top gate, H$_2$/Ar treatment, or current annealing was used.
The intrinsic electron-doping restricted us only to the electron-doped region for detailed $1/f$-noise measurements. $S_\sigma/\sigma^2$ in BLG
device clearly behaves very differently from the SLG devices in Fig.~2. At low $T$ ($\lesssim 150$~K), $S_\sigma/\sigma^2$ increases
monotonically with $|n|$, in agreement to our earlier results~\cite{atin PRL}, but becomes nonmonotonic at higher $T$, similar to the results
reported by Heller {\it et al.}~\cite{Charge noise}. Eq.~\ref{eq2} can however be fitted rather well to the BLG noise data over the entire range
of $n$ albeit with a positive $\gamma$ for all $T$. We find $\gamma \approx 1$ up to $T \sim 250~K$  (Fig.~4a). We suggest positive magnitude of
$\gamma$ to be due to the parabolic energy bands in BLG. Quantitatively, within the Thomas-Fermi approximation, $v_q$ is independent of $n$ for
BLG to the leading order~\cite{sdsarma_BLG}, resulting in $N_{con} \sim \ell^2|v_q|^4 \sim |n|$, {\it i.e.} $\gamma \approx 1$. Close to room
temperature, $\ell$ varies weakly with $n$, possibly due to competing scattering from the longitudinal acoustic
phonons~\cite{phonon_natnano,{phonon_avouris}}, causing the nonmonotonicity of $(d\sigma/dn)^2$ to be visible in the $n$-dependence of total
noise. We have also measured flicker noise in few-layer (FLG, $\sim 3 - 4$ layers, Fig.~3d) and many-layer (MLG, $\sim 15$ layer, Fig.~3e)
GraFET devices as well. Eq.~\ref{eq2} provides good fit to noise in these systems with a small positive $\gamma$ arising from the parabolic
bands at low energies.

\begin{figure}[!b]
\begin{center}
\includegraphics [width=1\linewidth]{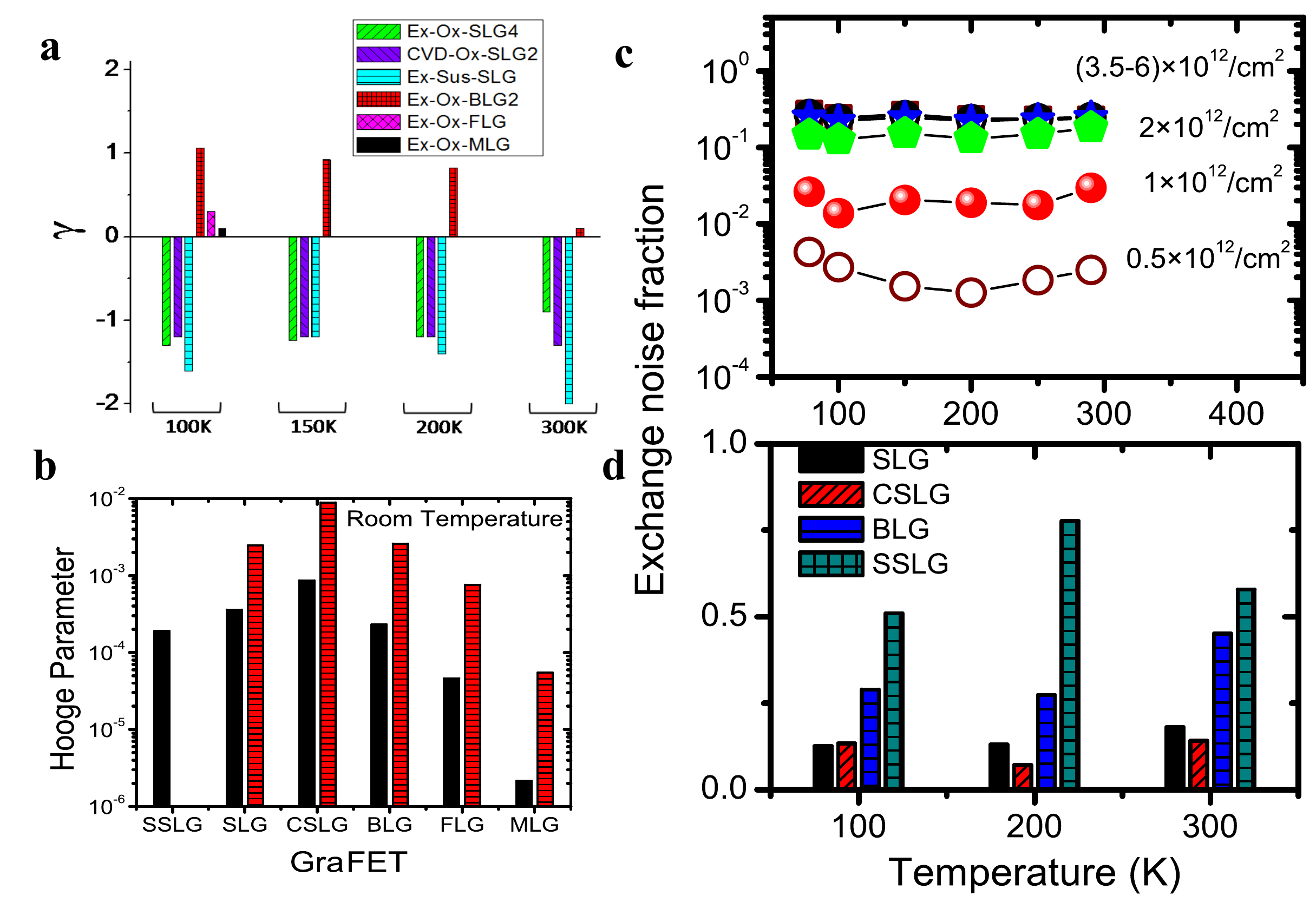}
\end{center}
\caption{(a) $\gamma$ (see text) at different temperatures, for devices used in this work. (b) Comparison of Hooge parameter ($\gamma_H$) for
various Gra-FET devices at room temperature: Suspended single layer graphene (SSLG), substrated single layer (SLG), bilayer (BLG), few layer
(FLG), many layer (MLG) and CVD-grown single layer graphene (CSLG). The solid black bar and dashed red bar correspond to $\gamma_H$ measured at
densities $n=2 \times 10^{11}$~/cm$^2$ and $n=2.4 \times 10^{12}$ /cm$^2$ respectively. (c) The exchange noise fractions are plotted with
temperature at different carrier densities ($n$) for the SLG device used in Fig.~2a1-a3. (d) The exchange noise fractions for four different
kind of GraFETs at three different temperatures at a carrier density $n=2n_i$ (see text). } \label{figure4}
\end{figure}

In essence, Graphene displays both trapping-detrapping-like noise in MOSFET
(exchange noise) and that from changes in extended structural disorder as in
disordered metal films (configuration noise). The origin of both in this case
is a fluctuating charge distribution, where the sensitivity of configurational
component (more specifically, $\gamma$) to band structure allows one to
distinguish between the linear and parabolic bands(Fig.~4a). In fact, we find
the configurational component to dominate in most devices and densities, as
indicated in Fig.~4c and 4d, but strongly depends on the quality of the
substrate surface, roughness, nature of dielectric, operating $n$ etc. This can
vary widely from one device to the other, helping us to understand the
apparently different experimental results on GraFET noise reported from
different research groups~\cite{Avouris 1,{atin PRL},{atin APL},{Charge
noise},{Noise_Xu},{Noise_Balandin},{Noise_suspended}}. Not surprisingly, the
exchange noise is maximum in the suspended devices (Fig.~4d), where the
discontinuous layer of surface residues leaves very little room for the trapped
charges to redistribute.

Fig.~4b summarizes normalized noise levels in different designs of GraFET. The comparison is made in terms of the phenomenological Hooge
parameter $\gamma_H$, defined as $\gamma_H = n(fS_\sigma)a_G/\sigma^2$ where $a_G$ is the area of the graphene film between voltage leads. At
all $n$, the substrated SLG devices, exfoliated or CVD-grown, are most noisy, whereas suspended SLG and thicker graphene systems are nearly
hundred times quieter. At room temperature, and even on a substrate, $\gamma_H$ can be $\sim 10^{-7} - 10^{-6}$ in FLG and MLG
devices~\cite{atin APL}, which is among the lowest known for metal or semiconductor nanostructures. The extreme low noise in these systems is
due to strong screening by the lower layers, which also affects the gating ability, limiting their usefulness in active electronics, but make
them suitable as interconnects.

\textbf{Acknowledgement} We acknowledge the Department of Science and Technology (DST) for a funded project. ANP, SG and VK thank CSIR for
financial support.

\end{document}